# Agile Software Effort Estimation using Regression Techniques

Sisay Deresa Sima
*Department of Computer Science*
*Addis Ababa University*
Addis Ababa,Ethiopia
sisay.deresa@aau.edu.et

Ayalew Belay Habtie
*Department of Computer Science*
*Addis Ababa University*
Addis Ababa,Ethiopia
ayalew.belay@aau.edu.et

*Abstract—* Software development effort estimation is one of the most critical aspect in software development process, as the success or failure of the entire project depends on the accuracy of estimations. Researchers are still conducting studies on agile effort estimation. The aim of this research is to develop a story point based agile effort estimation model using LASSO and Elastic Net regression techniques. The experimental work is applied to the agile story point approach using 21 software projects collected from six firms. The two algorithms are trained using their default parameters and tuned grid search with 5-fold cross-validation to get an enhanced model. The experiment result shows LASSO regression achieved better predictive performance PRED (8%) and PRED (25%) results of 100.0, MMRE of 0.0491, MMER of 0.0551, MdMRE of 0.0593, MdMER of 0.063, and MSE of 0.0007. The results are also compared with other related literature.

## I. INTRODUCTION

The software development paradigm moves to the agile development of software launched in 2001 [1]. The success rate has been enhanced for projects that use agile techniques [2]. The research [3] claims that 70% of the software for applications was produced by agile techniques in 2018. Agile techniques have become quite popular due to their capacity to adapt to the changing processes of software projects. Ahead of software development, effort estimates are the most crucial phases [2]. Estimating efforts to build software is the most challenging [4] technique for developing software because of its inaccurate, imprecise, and inadequate input [5]. Software estimates include [6]: an ideal time (productive project completion time), velocity (measuring overall user stories and executing tasks in the iteration cycle) and total effort (actual effort to finish the work within the entire period for the product to be completed).

Estimating software effort is essential to a successful software project since it impacts cost, time, quality, and performance [7]. Accurately estimating the effort leads to the project's success; if not, the project will fail [3]. The major reasons for software failure are poor management, preliminary estimates, and constant changes in needs, incomplete requirements, poor communication amongst developers, and the inability to identify risks early on [3]. Estimation is crucial in development, but no specific approach to estimating that software can be as accurate as possible [8].

Methodology of software effort estimation is primarily categorized into two, i.e., algorithmic and non-algorithmic [9]. Algorithmic estimates models are parametric and predicted with a few formulae set to historical data [10]. Non-algorithmic approaches decide and make conclusions depending on the data available and earlier project experience [8, 11]. Machine learning approaches are used to increase the accuracy of both algorithmic and non-algorithmic [12].

In agile, story point [13] is the most commonly used measuring unit to estimate effort. A user's story is essentially a user requirement [14], and it applies to all agile methods [15] effort estimating approach. The size and complexity of user stories are determined and assigned [16].Although researchers have applied various machine learning techniques for effort estimation using the story point approach, there is still a gap in estimation accuracy.

This study develops a story point agile effort estimation model using a publicly available dataset that applies LASSO and Elastic Net regression algorithms.

The organization of this paper is as follows: Section II contains related work, Section III pproposed model, Section IV implementation , Section V evaluation criteria,Section VI experiments and rresults and Section VII conclusion.

## II. RELATED WORK

Zia *et al.* [16] developed a model based on user stories for agile software development. The model was validated using twenty-one projects. The experimental result was evaluated using the mean magnitude relative error and prediction accuracy. This result implies that MMRE has met the acceptance criteria. However, the PRED(25) result is less than 0.75 [17] and failed Conte *et al.* [18] acceptance criteria.

Vyas and Hemrajani [19] proposed linear, ridge, and logistic regression models to estimate software development effort. Following steps were employed: Normalization, dataset splitting 80% for training and 25% for testing, and tuning parameters through the random search. Finally, they compared it with Zia's [16].It is observed that ridge regression revealed the highest value in terms of PRED(25) and MMRE. Though the result is acceptable, it still needs improvement for a better estimation accuracy.

Sharma and Chaudhary [20] computed the correlation between the dependent and independent variables. According to the correlation values, the effort is more associated with development time with the most significant correlation values. Three multiple linear regression models were made for effort estimation in agile software development. Among the three models, two of them were linear regression, and one was polynomial regression. The Squared Correlation Coefficient ($R^2$) and Mean Square Error (MSE) were used to measure the model's estimation accuracy. The linear regression model with an initial velocity, total velocity, and declaration factor outperforms Satapathy *et al.* [21] in terms of MMRE. The result is better and fulfills the threshold MMRE. Nevertheless, it is difficult to conclude on the acceptance because due to the sole MMRE result.

Arora *et al*. [22] used six regression techniques, namely: Extreme Gradient Boosting (XGB), Decision Tree (DT), Linear Regressor (LR), Random Forest (RF), Adaptive Boosting (AdaBoost), and Categorical Boosting (CatBoost) to estimate the effort of agile projects. The techniques random and grid searches were used to tune the models. The experimental result shows CatBoost Regressor resulted in the lowest root mean square error. Due to the sole result of prediction accuracy, it is not enough to conclude the models acceptance.

Satapathy *et al*. [23] applied Support Vector Regression (SVR) to enhance agile effort estimation. Four SVR kernels, namely, Sigmoid Kernel, Linear Kernel, Radial Basis Function (RBF) Kernel, and Polynomial Kernel, are used to implement the SVR-based model. The inputs of the model consist of total story points and project velocity from 21 software projects. The training set 80%, and the testing set 20%. The mean magnitude relative error and prediction accuracy are used to measure the models' estimation performance. The empirical result shows that SVR with RBF kernel achieved the highest performance and met the criteria set by Conte *et al*. [18]. However, Zakrani *et al*. [17] showed limitations in estimation accuracy.

Satapathy and Rath [21] utilized a story point method to reduce the number of errors in the accuracy by applying three distinct algorithms: Stochastic Gradient Boosting (SGB), Random Forest (RF), and Decision Tree (DT). The complete dataset was divided into training and testing sets for SGB and RF techniques. The training dataset accounted for 80% of the total, and the test dataset accounted for 20%. The SGB strategy employs tenfold cross-validation, whereas the RF technique uses leave-one-out validation. The results showed that the stochastic gradient boosting technique outperforms. However, MMRE is not stated, and it is challenging to decide whether the model is acceptable.

Zakrani *et al*. [17] proposed an improved effort estimation model for agile projects using Support Vector Regression (SVR) optimized by the grid search(GS). Five-fold cross-validation and Leave-one-out cross-validation (LOOCV) was used in the training and test phase, respectively. The comparison was performed using 21 agile software projects and three accuracy measures (MMRE, MdMRE, and PRED(0.25)) and compared with [16,21,23,24,25]. The result showed SVR-GS outperforming and satisfying the threshold. However, they do not mention the training and testing split ratio. In addition, the error value MMRE should be decreased for enhanced estimation of effort.

Saini *et al*. [15] developed a fuzzy approach based on the Mamdani Fuzzy Inference System to find agile software development efforts. This model used three input variables: user story, team expertise, and complexity on 21 projects dataset. However, the proposed model was not empirically evaluated.

Kaushik *et al.* [25] proposed fuzzy logic based agile software and cost estimation. They found 21 factors that affect the agile software development environment. These cost drivers were combined with the story point approach to estimate effort using fuzzy type-2, type-1 interval and compared with [16]. The outcome shows, a type-2 interval is outperforming and acceptable in terms of PRED(7.19). However, the error MMRE result was higher and failed to meet the threshold.

Panda *et al*. [26] compared different types of Neural Networks (NN) such as Probabilistic Neural Network (PNN), Group Method of Data Handling (GMDH) Polynomial Neural Network, General Regression Neural Network (GRNN), Cascade Correlation Neural Network (CCNN),and Zia's model [16] to improve the accuracy of agile effort. MMRE, PRED (25), MSE, $R^2$ metrics were used for empirical evaluation. The result showed that, except MMRE of GRNN, other neural network models met the threshold. However still, the models have limitations in estimation accuracy.

Bilgaiyan *et al*. [27] predicted software effort in agile software development (ASD) using Artificial Neural Network (ANN) trained using Back Propagation and Elman Neural Network (ENN).The result was also compared with [26]. The evaluation criteria were MSE, MMRE, and PRED. The effort estimation of the Feedforward Back-Propagation Network has been far better than Elman Neural Network and Cascade Correlation Neural Network. The empirical result shows, the MMRE did not satisfy the thresholds proposed by Conte *et al*. [18]; in contrast, the PRED (25%) is greater than the threshold set.

Rao *et al*. [3] proposed Adaptive Radial Basis Function Networks, Neuro-Fuzzy and Generalized Regression Neural Networks. Though, the models did not state the training and test ratio. The finding showed different results in terms of MMRE and MMER, but all produced approximately the same PRED, and none of the models satisfy Conte *et al*. [18] threshold.

Khuat and Le [24] aimed to ameliorate effort estimation by combining Particle Swarm Optimization (PSO) and Artificial Bee Colony (ABC) algorithms. MMRE, and PRED were considered performance metrics. A comparison has been made with Panda *et al*. [26] and Zia's regression [16]. However, the limitation of the study is they used all the data for training and testing [17].

Khuat and Le [28] presented an Artificial Neural Network (ANN) optimized by the combination of Fireworks and Levenberg-Marquardt algorithms to improve the estimation accuracy of Zia's [16] used fifteen projects for the training and six projects for testing and compared with [16, 25]. The experimental shows, they employed incorrect input effort while testing.

Sharma and Chaudhary [12] used agile and traditional methods. Neural Network methods with Backpropagation and Genetic Algorithms were employed. The input included velocity and user stories for an agile method, the kilo lines of code in the traditional method. The dataset is separated into training, validation, and testing, respectively. For agile, the dataset is divided into 70%, 15%, and 15% and for the traditional dataset is divided into 60%, 20%, and 20%. However, in both models, the value of the MMRE obtained is not less than 0.25 and did not meet the thresholds suggested by Conte *et al*. [18].On the other hand, the PRED (25) is not specified, which makes it difficult to decide the acceptance of the models.

Bilgaiyan *et al*. [4] developed Chaotically Modified Genetic Algorithm (CMGA) to address software development effort estimation (SDEE). MMRE and PRED (25) were considered as performance evaluation metrics. The CMGA method has been utilized to improve and optimize the settings of the Dilation-Erosion Perceptron (DEP), thereby enhancing the SDEE. Dataset was divided into three distinct sets where training set includes 50% of data, validation set includes 25% of data and test set contains 25% of data. The estimation results compared with [25, 26]. The proposed model exceeds and satisfies the thresholds but still needs to improve on estimation accuracy.

Kaushik *et al*. [1] investigated Deep Belief Network (DBN) and Antlion Optimization (ALO) techniques for effort prediction in agile and non-agile software development. Firefly-synthetic minority oversampling technique is applied for the agile dataset. Four non-agile datasets were taken from the PROMISE Software Engineering Repository. They statistically confirmed using Friedman and Wilcoxon signed-rank test. The experiments showed that DBN-ALO meets the threshold. However, Kaushik *et al*. [1] stated that actual estimation of effort does not require oversampling technique on agile datasets. Firefly-synthetic minority oversampling makes it difficult to generalize, unlike the real dataset.

Kaushik *et al.* [13] designed an effort estimation technique for agile software development using Radial Basis Function Neural Network (RBFN) and Functional Link Artificial Neural Network (FLANN),with Whale Optimization Algorithm (WOA). The proposed methods are evaluated on three agile datasets and used MMRE, MdMRE, and PRED (0.25) as evaluation metrics. The process was further validated using Friedman and Wilcoxon statistical tests. The experimentation found FLANN-WOA and RBFN-WOA failed to meet MMRE, PRED (0.25) dataset [16]. Though, FLANN-WOA and RBFN-WOA met the thresholds proposed by Conte *et al.* [18] the models still need enhancement in estimation accuracy.

There has been a lot of work in agile effort estimation [4, 13,22,23,25,29] that showed acceptable estimation metrics set by Conte *et al*. [18]. However, these studies still have limitations in estimation accuracy. Zia *et al*. [16] and Rao *et al.* [3] models failed to meet the acceptance threshold. Estimation results are also difficult to decide on the model acceptance because of the sole values in MMRE [12, 17, 20] and PRED(25%)[18][27]. Zakrani *et al*. [17] have MMRE that needs to be minimized for enhancement of prediction and its train test split ratio is not also stated. Khut and Le [24], used the same dataset for training and testing [17] rather than splitting into training and test set and used incorrect input data while testing in their other study [28].

Because of the above reasons we proposed an enhanced agile effort estimation model. Here,we leverage estimation capabilities of both Elastic Net and LASSO regression to design an agile effort estimation model from [30] and Grid Search [17] using a dataset available by Zia *et al.* [16].

### III. PROPOSED MODEL

We proposed agile effort estimation model using Zia *et al.* [16] dataset, which consists of 21 projects developed by six software houses. Our models use the total count of story points and project velocity as input parameters. The sklearn provides libraries of LASSO and Elastic Net regressions algorithms. The proposed approach includes data collection, normalization, train-test split, training with default parameters, training with cross-validation and evaluation. The following phases are applied to build the models:

A. *Data Collection*: the total count of story points and project velocity are considered input parameters taken from [16].

B. *Normalization:* it deals with producing the value of effort, velocity, and time within the range of [0,1].

C. *Train-test Split:* we divided the dataset into training 80% and 20% testing.

D. *Training with Default Parameters:* Elastic Net and LASSO regression default parameters stated in Table 1 used during training model. Finally, we evaluate its performance using test data to produce estimation results.

E. *Training with Cross-Validation:* grid search with fivefold cross-validation used to develop final model, then it is evaluated on the testing set to produce the final enhanced estimation result.

F. *Evaluation:* to evaluate the estimation of effort for agile effort metrics such as PRED (8%), PRED (25%), MMRE, MMER, MdMRE, MdMER, and MSE are used.

Table 1 Default and Optimal Parameters

| Techniques | Parameters | Default values | Best values |
|---|---|---|---|
| Elastic Net regression | alpha | 1.0 | 0.001 |
| | l1_ratio | 0.5 | 0.001 |
| | max_iter | 1000 | 25 |
| | random_state | None | 120 |
| LASSO regression | alpha | 1.0 | 0.001 |
| | max_iter | 1000 | 25 |
| | random_state | None | 120 |

### IV. IMPLEMENTATION

A. *LASSO Regression*

The least absolute shrinkage and selection operator (LASSO) is the same as ridge regression [31]. LASSO is capable of compensating for the absolute magnitude of regression coefficients as well as reducing inconsistency and improving the quality of linear regression models. The LASSO regression equation varies from the ridge regression equation in that it utilizes the absolute value of the coefficients rather than the squares as indicated in Equation 1.

$$MIN \| (X_w - y)^2 \| + \delta \|w\| \quad (1)$$

where $MIN \| (X_w - y)^2 \|$ is the least square expression, $\delta$ is shrinkage parameter, *w* is the coefficient.

B. *Elastic Net Regression*

Elastic net regression is favored over ridge and lasso regression when dealing with highly correlated independent variables [31]. It is a combination of both LASSO and ridge regression shown in Equation 2.

$$MIN \| (X_w - y)^2 \| + \delta_1 \|w^2\| + \delta_2 \|w\| \quad (2)$$

Elastic net regression is useful when there are correlated multiple features. In the case of LASSO, one of these is chosen, but in the elastic net, both are chosen.

### V. EVALUATION CRITERIA

- **Mean Squared Error (MSE):** is computed by Equation 3. The lower the values of MSE, the more accurate the values of estimated result.

$$\mathbf{MSE} = \frac{1}{N} \sum_{i=1}^{N} (AE_i - EE_i)^2 \quad (3)$$

where N is the total number of observations, $AE_i$ and $EE_i$ are actual and estimated effort of the $i^{th}$ test data respectively.

- **Magnitude of Relative Error (MRE):** calculated for each project in the dataset, while mean magnitude of relative error (MMRE) computes the average over N observations as illustrated in Equation 4.

$$MRE_i = \frac{|AE_i - EE_i|}{|A_i|} \quad (4)$$

where $AE_i$ is actual effort and $EE_i$ estimated effort of the $i^{th}$ test data respectively.

- **Mean Magnitude of Relative Error (MMRE):** measures the difference between estimated and actual value relative to actual value, as given in Equation 5.

$$MMRE = \frac{1}{N}\sum_{i=1}^{N} MRE_i \quad (5)$$

- **Prediction Accuracy (PRED):** is defined in Equation 6.

$$Pred(N) = \frac{100}{N}\sum_{i=1}^{N}\begin{cases}1, & if\ MRE_i \leq \frac{N}{100}\\0, & otherwise\end{cases} \quad (6)$$

where N is the total number of data in the test set

- **Median Magnitude of Relative Error (MdMRE):** equation 7 shows how to calculate MdMRE.

$$MdMRE = median(MRE_i) \quad (7)$$

The accuracy of a suggested model for the MMRE or MdMRE is higher when the metric result is closer to zero.

- **The squared correlation coefficient ($R^2$):** measures the relationship between the experimental data and the predicted data, and the algorithm has $R^2$ value closer to 1 and defined in Equation 8.

$$R^2 = 1 - \left(\frac{\sum_{i=1}^{N}(AE_i - EE_i)^2}{\sum_{i=1}^{N}(AE_i - AE_{mean})^2}\right) \quad (8)$$

where $AE_{mean}$ is the mean of actual effort.

- **Root mean squared (RMSE):** calculated as Equation 9.

$$RMSE = \sqrt{\sum_{i=1}^{N}\frac{(AE_i - EE_i)^2}{N}} \quad (9)$$

## VI. EXPERIMENTS AND RESULTS

The estimation performance of the model was evaluated using mainly MMRE and PRED (25%) metrics. These metrics determine whether the proposed model is acceptable or not in different researches.

### A. Comaparion of Elastic Net and LASSO regression

Table 2 shows both algorithms yield almost same result in default parameters. However, Elastic Net regression is slightly higher in MMER and MdMER (+0.001 and +0.001), respectively. Therefore, LASSO with tuning is a more acceptable model, and will be using the LASSO with tuning model for comparison with other researches.

Table 2 Comaparion of Elastic Net and LASSO regression

| Technique | MMRE | MMER | MdMRE | MdMER | PRED (8%) | PRED (25%) |
|---|---|---|---|---|---|---|
| Elastic Net with default parameters | 0.7193 | 0.7468 | 0.1592 | 0.1373 | 40 | 60 |
| LASSO with default parameters | 0.7193 | 0.7468 | 0.1592 | 0.1373 | 40 | 60 |
| Elastic Net with Tuning | 0.0490 | 0.0547 | 0.0574 | 0.0609 | 100 | 100 |
| LASSO with Tuning | 0.0490 | 0.0546 | 0.0570 | 0.0604 | 100 | 100 |

### B. Comaparion with Related Researches

Further comparison is made with other related researches and grouped by similar metrics. The comparison result in Table 3 shows that the LASSO with tuning regression model attained the highest estimation accuracy and acceptable than the methods [3,4,19,23,25,26].

Table 3 Comparison with MMRE and Prediction

| References | MMRE | PRED(0.25) |
|---|---|---|
| Satapathy et al. [23] | 0.0747 | 95.9052 |
| Panda et al.[26] | 0.3581 | 85.9182 |
|  | 1.5776 | 87.6561 |
|  | 0.1563 | 89.6689 |
|  | 0.1486 | 94.7649 |
| Rao et al. [3] | 8.4277 | 76.1905 |
|  | 2.7864 | 76.1905 |
|  | 8.0909 | 76.1905 |
|  | 6.6430 | 76.1905 |
| Bilgaiyan et al. [4] | 0.0565 | 96.79 |
| Kaushik et al. [25] | 46.43 | 14.28 |
|  | 4.20 | 90.47 |
| Vyas and Hemrajani [19] | 0.15 | 71.42 |
|  | 0.19 | 71.42 |
|  | 0.13 | 85.71 |
| LASSO with Tuning | 0.0490 | 100 |

Table 4 shows LASSO with Tuning model estimation accuracy comparison with [12][20][22] regarding PRED (25), $R^2$, MSE, and RMSE metrics. LASSO with Tuning model outperforms in all metrics. However, in Table 4, it is difficult to conclude on the other models' acceptance with the absences of MMRE results.

Table 4 Comparison of Results

| References | PRED (25) | $R^2$ | MSE | RMSE |
|---|---|---|---|---|
| Sharma et al. [20] | NA | 0.9476 | 718.1487 | NA |
| Sharma et al. [12] | NA | 0.973897 | 17.03561 | NA |
|  |  | 0.967112 | 21.46326 |  |
| Arora et al. [22] | 0.9345 | 0.93 | 34.25 | 5.8523 |
|  | 0.9215 | 0.92 | 41.05 | 6.4070 |
|  | 0.9179 | 0.9179 | 42.991 | 6.5505 |
| LASSO with Tuning | 100 | 0.9760 | 0.0007 | 0.1760 |

Table 5 Comparison of Results

| References | MSE | MMRE | MMER | MdMRE | MdMER | PRED (8%) | PRED (25%) |
|---|---|---|---|---|---|---|---|
| Zia et al. [16] | NA | 0.0719 | NA | 0.0714 | NA | NA | 57.14 |
| Satapathy et al. [21] | NA | NA | 0.3820 | NA | 0.2896 | NA | 38.0952 |
|  |  |  | 0.1632 |  | 0.1151 |  | 85.7143 |
|  |  |  | 0.2516 |  | 0.2033 |  | 66.6667 |
| Zakrani et al. [17] | NA | 0.0620 | 0.0613 | 0.0426 | 0.0408 | 66.667 | 100 |
| Bilgaiyan et al. [27] | 0.056 | 0.1480 | NA | NA | NA | NA | 94.8659 |
|  | 0.052 | 0.1349 |  |  |  |  | 95.2301 |
| Kaushik et al.[13] | NA | 0.186 | NA | 1.73 | NA | NA | 86.4 |
|  |  | 0.198 |  | 1.73 |  |  | 95.2301 |
| Kaushik et al. [1] | NA | 0.0225 | NA | 0.0222 | NA | NA | 98.4321 |
| LASSO with Tuning | 0.0007 | 0.0490 | 0.0546 | 0.0570 | 0.0604 | 100 | 100 |

Table 5 shows estimation accuracy comparison [1,13,16, 17,21,27] in MMRE and PRED(25%) other metrics. The experimental result shows we achieved the lowest error MSE and highest PRED(8%).However, MdMRE and MdMER are slightly higher (+0.0144 and +0.0196), respectively, to Zakrani et al. [17]. Because the error metrics MMRE is less than [17] indicates less estimation error. Therefore, we can conclude that our model is more acceptable.

## VII. CONCLUSION

Effort estimation in agile development is still an issue. Machine learning techniques are widely used to enhance effort estimation on the story point approach. However, there is a limitation in the prediction accuracy of the models proposed by different researchers.

This research proposed a model that estimates effort using a story point approach applying Elastic Net and LASSO regression. Two scenarios are used: first, models are trained with their default parameters and assessed on a different test set in the first experiment. In the second scenario, five-fold cross-validation is applied for both algorithms using grid-search techniques to determine the optimum value and improve the model.

The results show that LASSO regression has acquired the result PRED(8%) and PRED (25%) result 100.0, MMRE 0.0490, MMER of 0.0546, MdMRE of 0.0570, MdMER of 0.0604 and MSE of 0.0007. Therefore, the result shows, it enhanced all metrics except slightly higher MdMRE and MdMER compared to Zakrani et al. [17].

REFERENCES

[1] A. Kaushik, D. K. Tayal, and K. Yadav, "A Comparative Analysis on Effort Estimation for Agile and Non-agile Software Projects Using DBN-ALO," *Arab. J. Sci. Eng.*, vol. 45, no. 4, pp. 2605–2618, 2020, doi: 10.1007/s13369-019-04250-6.

[2] M. S. Khan, C. A. U Hassan, M. A. Shah, and A. Shamim, "Software cost and effort estimation using a new optimization algorithm inspired by strawberry plant," *ICAC 2018 - 2018 24th IEEE Int. Conf. Autom. Comput. Improv. Product. through Autom. Comput.*, pp.1–6,2018,doi: 10.23919/IConAC.2018.8749003.

[3] C. Prasada Rao, P. Siva Kumar, S. Rama Sree, and J. Devi, *An agile effort estimation based on story points using machine learning techniques*, vol. 712. Springer Singapore, 2018.

[4] S. Bilgaiyan, K. Aditya, S. Mishra, and M. Das, "Chaos-based modified morphological genetic algorithm for software development cost estimation," *Adv. Intell. Syst. Comput.*, vol. 710, pp. 31–40, 2018, doi: 10.1007/978-981-10-7871-2_4.

[5] C. Ratke, H. H. Hoffmann, T. Gaspar, and P. E. Floriani, "Effort Estimation using Bayesian Networks for Agile Development," *2nd Int. Conf. Comput. Appl. Inf. Secur. ICCAIS 2019*, pp. 1–4, 2019, doi: 10.1109/CAIS.2019.8769455.

[6] S. Dhir, D. Kumar, and V. B. Singh, "Feedforward and Feedbackward Approach-Based Estimation Model for Agile Software Development," pp. 73–80, 2017, doi: 10.1007/978-981-10-3770-2.

[7] T. E. Ayyıldız and H. C. Terzi, "Case Study on Software Effort Estimation," *Int. J. Inf. Electron. Eng.*, vol. 7, no. 3, pp. 103–107, 2017, doi: 10.18178/ijiee.2017.7.3.670.

[8] R. Popli, "ICROIT 2014 - Proceedings of the 2014 International Conference on Reliability, Optimization and Information Technology," *ICROIT 2014 - Proc. 2014 Int. Conf. Reliab. Optim. Inf. Technol.*, pp. 57–61, 2014.

[9] R. R. Sinha and R. K. Gora, "Software effort estimation using machine learning techniques," *Lect. Notes Networks Syst.*, vol. 135, pp. 65–79, 2021, doi: 10.1007/978-981-15-5421-6_8.

[10] P. Suresh Kumar and H. S. Behera, "Role of Soft Computing Techniques in Software Effort Estimation: An Analytical Study," *Adv. Intell. Syst. Comput.*, vol. 999, pp. 807–831, 2020, doi: 10.1007/978-981-13-9042-5_70.

[11] R. K and Beena R, "A Critique on Software Cost Estimation," *Int. J. Pure Appl. Math.*, vol. 118, no. 20, pp. 3851–3862, 2018.

[12] A. Sharma and N. Chaudhary, "Analysis of Software

Effort Estimation Based on Story Point and Lines of Code using Machine Learning," *Int. J. Comput. Digit. Syst.*, pp. 1–8, 2021, [Online]. Available: https://journal.uob.edu.bh:443/handle/123456789/4491.

[13] A. Kaushik, D. K. Tayal, and K. Yadav, "The role of neural networks and metaheuristics in agile software development effort estimation," *Int. J. Inf. Technol. Proj. Manag.*, vol. 11, no. 2, pp. 50–71, 2020, doi: 10.4018/IJITPM.2020040104.

[14] M. Vyas, "A review on software cost and effort estimation techniques for agile development," vol. 5, no. 1, pp. 1–5, 2020.

[15] A. Saini, L. Ahuja, and S. K. Khatri, "Effort Estimation of Agile Development using Fuzzy Logic," *2018 7th Int. Conf. Reliab. Infocom Technol. Optim. Trends Futur. Dir. ICRITO 2018*, pp. 779–783, 2018, doi: 10.1109/ICRITO.2018.8748381.

[16] S. Kamal Tipu and S. Zia, "An Effort Estimation Model for Agile Software Development," *Adv. Comput. Sci. its Appl.*, vol. 314, no. 1, pp. 2166–2924, 2012, [Online]. Available: www.worldsciencepublisher.org.

[17] A. Zakrani, A. Najm, and A. Marzak, "Support Vector Regression Based on Grid-Search Method for Agile Software Effort Prediction," *Colloq. Inf. Sci. Technol. Cist*, vol. 2018-Octob, pp. 492–497, 2018, doi: 10.1109/CIST.2018.8596370.

[18] V. Y. Conte, S. D., Dunsmore, H. E., & Shen, *"Software Engineering Metrics and Models."* Redwood City, CA, USA: Benjamin-Cummings, 1986.

[19] M. Vyas and N. Hemrajani, "Predicting Effort of Agile Software Projects using Linear Regression, Ridge Regression and Logistic Regression," *Int. J. "Technical Phys. Probl. Eng.*, vol. 13, no. 47, pp. 14–19, 2021.

[20] A. Sharma and N. Chaudhary, "Linear Regression Model for Agile Software Development Effort Estimation," *2020 5th IEEE Int. Conf. Recent Adv. Innov. Eng. ICRAIE 2020 - Proceeding*, vol. 2020, pp. 4–7, 2020, doi: 10.1109/ICRAIE51050.2020.9358309.

[21] S. M. Satapathy and S. K. Rath, "Empirical assessment of machine learning models for agile software development effort estimation using story points," *Innov. Syst. Softw. Eng.*, vol. 13, no. 2–3, pp. 191–200, 2017, doi: 10.1007/s11334-017-0288-z.

[22] M. Arora, M. Malviya, A. Sharma, S. Chopra, and S. Katoch, "A State of the Art Regressor Model's comparison for Effort Estimation of Agile software," pp. 211–215, 2021.

[23] S. M. Satapathy, A. Panda, and S. K. Rath, "Story point approach based agile software effort estimation using various SVR kernel methods," *Proc. Int. Conf. Softw. Eng. Knowl. Eng. SEKE*, vol. 2014-Janua, no. January, pp. 304–307, 2014.

[24] T. T. Khuat and M. H. Le, "A Novel Hybrid ABC-PSO Algorithm for Effort Estimation of Software Projects Using Agile Methodologies," *J. Intell. Syst.*, vol. 27, no. 3, pp. 489–506, 2018, doi: 10.1515/jisys-2016-0294.

[25] A. Kaushik, D. K. Tayal, and K. Yadav, "A Fuzzy Approach for Cost and Time Optimization in Agile Software Development," *Adv. Intell. Syst. Comput.*, vol. 1082, pp. 629–639, 2020, doi: 10.1007/978-981-15-1081-6_53.

[26] A. Panda, S. M. Satapathy, and S. K. Rath, "Empirical Validation of Neural Network Models for Agile Software Effort Estimation based on Story Points," *Procedia Comput. Sci.*, vol. 57, pp. 772–781, 2015, doi: 10.1016/j.procs.2015.07.474.

[27] S. Bilgaiyan, S. Mishra, and M. Das, "Effort estimation in agile software development using experimental validation of neural network models," *Int. J. Inf. Technol.*, vol. 11, no. 3, pp. 569–573, 2019, doi: 10.1007/s41870-018-0131-2.

[28] T. T. Khuat, M. H. Le, Tung Khuat, and Hanh Le, " An Effort Estimation Approach for Agile Software Development using Fireworks Algorithm Optimized Neural Network.," *Int. J. Comput. Sci. Inf. Secur.*, vol. 14, no. 7, pp. 122–130, 2016.

[29] B. Tanveer, L. Guzmán, and U. M. Engel, "Effort estimation in agile software development: Case study and improvement framework," *J. Softw. Evol. Process*, vol. 29, no. 11, pp. 1–14, 2017, doi: 10.1002/smr.1862.

[30] A. G. P. Varshini and K. A. Kumari, "Predictive analytics approaches for software effort estimation: A review," no. 2, pp. 2094–2103, 2020.

[31] S. Mensah, J. Keung, M. F. Bosu, and K. E. Bennin, "Duplex output software effort estimation model with self-guided interpretation," *Inf. Softw. Technol.*, vol. 94, pp. 1–13, 2018, doi: 10.1016/j.infsof.2017.09.010.